\documentclass[11pt,fleqn]{article}
\usepackage{amsmath,amssymb,graphicx}
\usepackage{color,soul}

\begin{document}
\sloppy

\newtheorem{axiom}{Axiom}[section]
\newtheorem{Xclaim}[axiom]{Claim}
\newtheorem{Xconjecture}[axiom]{Conjecture}
\newtheorem{Xcorollary}[axiom]{Corollary}
\newtheorem{Xdefinition}[axiom]{Definition}
\newtheorem{Xexample}[axiom]{Example}
\newtheorem{Xlemma}[axiom]{Lemma}
\newtheorem{Xobservation}[axiom]{Observation}
\newtheorem{Xopen}[axiom]{Open Problem}
\newtheorem{Xproposition}[axiom]{Proposition}
\newtheorem{Xtheorem}[axiom]{Theorem}

\newcommand{\xproof}{\emph{Proof.}\ \ }
\newcommand{\xqed}{~$\Box$}
\newcommand{\rz}{{\mathbb{R}}}
\newcommand{\nz}{{\mathbb{N}}}
\newcommand{\zz}{{\mathbb{Z}}}
\newcommand{\eps}{\varepsilon}
\newcommand{\cei}[1]{\lceil #1\rceil}
\newcommand{\flo}[1]{\left\lfloor #1\right\rfloor}
\newcommand{\seq}[1]{\langle #1\rangle}

\newcommand{\boxxx}[1]
 {\fbox{\begin{minipage}{13.00cm}#1\end{minipage}}}

\newcommand{\opena}{\mbox{$($$\ast$$)$}}
\newcommand{\openb}{\mbox{$($$\ast$$\ast$$)$}}
\newcommand{\openc}{\mbox{$($$\ast$$\ast$$\ast$$)$}}

\newcommand{\A}{\underline{A}}
\newcommand{\B}{\underline{B}}
\newcommand{\C}{\underline{C}}
\newcommand{\D}{\underline{D}}
\newcommand{\W}{\underline{W}}
\newcommand{\X}{\underline{X}}
\newcommand{\Y}{\underline{Y}}
\newcommand{\Z}{\underline{Z}}

\newcommand{\core}{{\sc Core-Sta\-bi\-lity}}
\newcommand{\num}{\mbox{num}}
\newcommand{\vmax}{v^{\max}}

\title{{\bf Core stability in hedonic coalition formation}}
\author{\sc Gerhard J.\ Woeginger\thanks{{\tt gwoegi@win.tue.nl}. 
Department of Mathematics and Computer Science, TU Eindhoven, P.O.\ Box 513, 5600 MB Eindhoven, Netherlands}}
\date{}
\maketitle

\begin{abstract}
In many economic, social and political situations individuals carry out activities in groups (coalitions) 
rather than alone and on their own. 
Examples range from households and sport clubs to research networks, political parties and trade unions. 
The underlying game theoretic framework is known as coalition formation. 

This survey discusses the notion of core stability in hedonic coalition formation (where each player's 
happiness only depends on the other members of his coalition but not on how the remaining players outside 
his coalition are grouped). 
We present the central concepts and algorithmic approaches in the area, provide many examples, and pose 
a number of open problems.

\medskip\noindent\emph{Keywords:}
computational social choice; computational complexity; coalition formation; hedonic game. 
\end{abstract}

\section{Introduction}
In economic, social and political situations individuals often carry out activities in groups (coalitions) 
rather than alone and on their own. 
Examples range from households, families and sport clubs to research networks, political parties and trade unions. 
The underlying game theoretic framework is known as \emph{coalition formation}.  
In \emph{hedonic coalition formation} each player's happiness/satisfaction only depends on the other members 
of his coalition, but not on how the remaining players outside his coalition are grouped together.
The study of coalition formation in hedonic games goes back to the seminal paper \cite{DrGr1980} of
Dr\`eze \& Greenberg.

A central question in coalition formation concerns the \emph{stability} of a system of coalitions: 
if there is a possibility of increasing one's happiness/satisfaction by moving to another coalition 
or by merging or splitting or otherwise restructuring coalitions, players will react accordingly and 
the system will become unstable.
The social choice literature knows a wide variety of stability concepts, as for instance the core, the strict core, 
the Nash stable set, the individually stable set, and the contractually individually stable set. 
A research line initiated by Banerjee, Konishi \& S\"onmez \cite{BaKoSo2001} and by Bogomolnaia \& Jackson 
\cite{BoJa2002} concentrates on sufficient conditions that guarantee the existence of such stable solutions.
Computational complexity issues related to hedonic coalition formation have first been investigated by 
Ballester \cite{Ballester2004} who establishes the NP-completeness of detecting core stable, Nash stable, 
and individually stable partitions (under appropriately chosen encodings of the input).

This paper zooms into \emph{core stability}, a particularly active subarea of hedonic coalition formation games.
We survey the central computational questions of this rich and colorful area, and we will see that their
algorithmic behavior is surprisingly diverse.
The underlying main definitions are introduced in Section~\ref{sec:defi}.
Each of the remaining sections discusses one particular type of hedonic game, summarizes the known results on
algorithms and complexity, provides examples, and also poses a number of open problems.
The open problems are marked in the following way: 
{\opena} marks a problem that should be doable;
{\openb} means that the problem might be difficult; 
{\openc} marks a hard and outstanding problem.

\section{Basic definitions and first observations}
\label{sec:defi}
\nopagebreak
Let $N$ be a finite set of players.
A \emph{coalition} is a non-empty subset of $N$.
Every player $i\in N$ ranks all the coalitions containing $i$ via his preference relation $\preceq_i$; this order
relation is reflexive ($S\preceq_i S$), transitive ($S\preceq_i T$ and $T\preceq_i U$ implies $S\preceq_i U$) and 
complete (at least one of $S\preceq_i T$ and $T\preceq_i S$ holds), but it is not necessarily anti-symmetric
(so that $S\preceq_i T$ and $T\preceq_i S$ may hold simultaneously).
The underlying strict order is denoted $\prec_i$, where $S\prec_iT$ means that $S\preceq_iT$ but not $T\preceq_iS$.
If $S\prec_iT$ then player $i$ \emph{prefers} participating in $T$ to participating in $S$, and
if $S\preceq_i T$ then player $i$ \emph{weakly prefers} participating in $T$ to participating in $S$.

A \emph{partition} $\Pi$ is simply a collection of coalitions which partitions $N$; hence every coalition in 
$\Pi$ is non-empty, distinct coalitions are disjoint, and the union of all coalitions equals $N$.
For a partition $\Pi$ and a player $i$, we denote by $\Pi(i)$ the unique coalition in $\Pi$ containing player $i$.
The following definition introduces core stability, the key concept of this paper.

\begin{Xdefinition}
\label{df:core}
A coalition $S$ \emph{blocks} a partition $\Pi$, if every player $i\in S$ strictly prefers $\Pi(i)\prec_i S$.
A partition $\Pi$ is \emph{core stable}, if there is no blocking coalition.
\end{Xdefinition}

Intuitively speaking, the players in a blocking coalition would like to separate and form their own coalition,
which makes the underlying partition unstable.
The game is hedonic, since the satisfaction/dissatisfaction of a player only depends on the other members of
his coalition, but not on the grouping of the remaining players outside his coalition.

A closely related stability notion is strict core stability.
A coalition $S$ \emph{weakly blocks} a partition $\Pi$, if every player $i\in S$ weakly prefers 
$\Pi(i)\preceq_i S$, and if at least one player $j\in S$ strictly prefers $\Pi(j)\prec_j S$.
A partition $\Pi$ is \emph{strictly core stable}, if it has no weakly blocking coalition.
Note that a strictly core stable partition is also core stable.
While our main focus in this survey is on core stability, we will from time to time also point out 
results on strict core stability.

\begin{Xexample}
\label{ex:3cycle}
Consider a situation with three players $a,b,c$ that have the following preferences over their coalitions:
\begin{quote}
Preferences of player $a$:~~ $ab>ac>a>abc$

Preferences of player $b$:~~ $bc>ab>b>abc$

Preferences of player $c$:~~ $ac>bc>c>abc$
\end{quote}
There are only five possible partitions of the players:
the partition $\{abc\}$ is blocked by $a$,
$\{ab,c\}$ is blocked by $bc$,
$\{ac,b\}$ is blocked by $ab$,
$\{bc,c\}$ is blocked by $ac$, and
$\{a,b,c\}$ is blocked by $ab$.
Hence there is no core stable partition, and there also is no strictly core stable partition.
\end{Xexample}

\begin{figure}[bth]
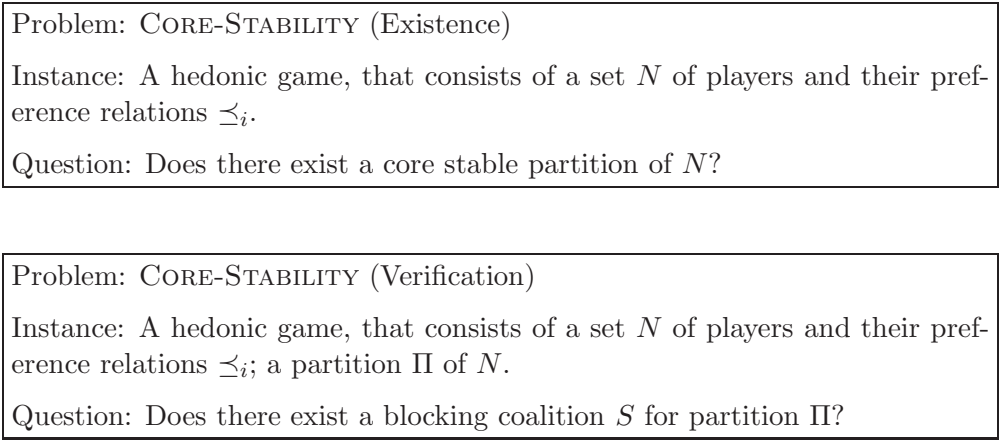

\begin{center}
\boxxx{Problem: {\core} (Existence)
\\[1.3ex]
Instance: A hedonic game, that consists of a set $N$ of players and their preference relations $\preceq_i$.
\\[1.3ex]
Question: Does there exist a core stable partition of $N$?}

\bigskip\bigskip
\boxxx{Problem: {\core} (Verification)
\\[1.3ex]
Instance: A hedonic game, that consists of a set $N$ of players and their preference relations $\preceq_i$;
a partition $\Pi$ of $N$.
\\[1.3ex]
Question: Does there exist a blocking coalition $S$ for partition $\Pi$?}
\end{center}
\caption{The algorithmic problems around core stability.}
\label{fig:problems}
\end{figure}

The central algorithmic questions are of course to decide whether a given game possesses a core stable partition
(existence problem) and to check whether a given partition for a given game actually is core stable
(verification problem).
Both problems are formally specified in Figure~\ref{fig:problems}.
The precise computational complexity of these two problems depends on the way the preference relations are 
specified in the input, and we will see a variety of natural ways in the following chapters.  
Throughout we only consider representations of the input for which the preference relations can be evaluated
in polynomial time: given a player $i$ and two coalitions $S$ and $T$ with $i\in S$ and $i\in T$, we can
decide in polynomial time whether $S\preceq_i T$.
Consequently we are also able to decide in polynomial time whether a given coalition $S$ blocks a given 
partition $\Pi$ of a given hedonic game.

Here is a rewording of the existence problem that clearly shows the quantifiers in the underlying question:
\begin{quote}
Does there exist a partition $\Pi$ of the players, such that every coalition $S$
satisfies the property that $S$ is not blocking for $\Pi$?
\end{quote}
Or even shorter: $\exists\Pi~\forall S$:$~\lnot$($S$ blocks $\Pi$).
This reworded formulation starts with an existential quantification, followed by an existential quantification, 
followed by a property that can be verified in polynomial time, and hence is a $\Sigma_2^p$-formulation; 
see for instance Theorem {17.8} in Papadimitriou \cite{PapaBook}.

\begin{Xobservation}
\label{ob:basic.1}
As the preference relations can be evaluated in polynomial time, the existence version of {\core}
is contained in the complexity class $\Sigma_2^p$ and the verification version is contained in NP.
\xqed
\end{Xobservation}

The verification problem $\exists S$:~($S$ blocks $\Pi$) is the negation of the inner part of the existence 
problem $\exists\Pi~\forall S$:$~\lnot$($S$ blocks $\Pi$). 
This yields the following straightforward connection between the two problems.

\begin{Xobservation}
\label{ob:basic.2}
If the verification problem of {\core} is polynomially solvable, then the existence version of {\core}
is contained in the complexity class NP.
\xqed
\end{Xobservation}

There are no further close connections between existence problem and verification problem.
In particular, hardness of the verification problem does not necessarily imply hardness of the existence problem:
under the enemy-oriented preferences in Section~\ref{sec:graphs} existence is easy whereas verification is hard.
We close this section with the common ranking property of Farrell \& Scotchmer \cite{FaSc1988} for
hedonic games in which all players have the same opinion about their coalitions.
Since such a game has little potential for disagreement, there always is a core stable partition:

\begin{Xobservation}
\label{ob:basic.3}
(Farrell \& Scotchmer \cite{FaSc1988})~
Consider a hedonic game for which there exists a function $f:2^N\to\rz$ such that $S\preceq_i T$ 
(for any player $i$ with $i\in S$ and $i\in T$) always implies $f(S)\le f(T)$.
Then this game has a core stable partition.
\end{Xobservation}
\xproof
Pick a coalition $S$ that maximizes the value of function $f$.
As the players in $S$ prefer $S$ to all other coalitions, they will never participate in a blocking coalition.
Repeat this step for the remaining players.
\xqed

\section{Complexity of the most general variants}
\label{sec:general}
\nopagebreak
We already noted earlier that the computational complexity of the {\core} existence and verification 
problem does heavily depend on the representation of the input.

The \emph{trivial encoding} of an $n$-player game presents the preference relation of every player 
by explicitly listing and ranking \emph{all} the $2^{n-1}$ coalitions that contain that player.
If we ignore polynomial factors, the resulting encoding length is roughly $L\approx2^n$.
As the verification problem can be solved by searching through all $2^n\approx L$ coalitions, it
is polynomially solvable under the trivial encoding.
The existence problem can be solved similarly by searching through all partitions of the player set. 
Now recall that an $n$-element set has roughly $c^{n\log n}$ different partitions (where $c$ is some 
real constant; see for instance De Bruijn \cite{Bruijn1958}), so that the resulting time complexity 
would be roughly proportional to $L^{\log\log L}$.
On the one hand this time complexity is bad, as it is not polynomially bounded in the encoding length $L$.
But on the other hand this time complexity is quite good, as it has a very mild and innocent sub-exponential 
growth rate; this makes it unlikely that the existence problem could be NP-complete.

\begin{Xopen}
\label{open:trivial}
{\openb}
Pinpoint the computational complexity of the {\core} existence problem under the trivial encoding.
\end{Xopen}

A good starting point for this open problem might perhaps be the complexity class LOGLOGNP introduced 
by Papadimitriou \& Yannakakis in \cite{PaYa1996}.
In any case the trivial encoding is inefficient, wasteful and unwieldy, and it definitely is not 
the right way for encoding preference structures in the real world.

A central assumption in cooperative game theory is \emph{individual rationality}, which says that no player 
should receive less than what he could get on his own.
In our framework a coalition $S$ is individually rational for player $i$ iff it satisfies $\{i\}\preceq_i S$.
If a coalition is not individually rational, it is blocked by some single-player set and hence can never occur
in a core stable partition.
In this spirit Ballester \cite{Ballester2004} suggests to specify the preference relation for player $i$ by 
listing only his individually rational coalitions; we call this the \emph{individually rational encoding} of 
the input.

\begin{Xtheorem}
\label{th:Ballester}
(Ballester \cite{Ballester2004})~
Under the individually rational encoding, the {\core} verification problem is polynomially solvable and
the {\core} existence problem is NP-complete.
\end{Xtheorem}
\xproof
The verification problem is straightforward, as the input explicitly lists all candidates for a blocking set.
By Observation~\ref{ob:basic.2} then the existence problem lies in NP, and it remains to establish its NP-hardness.

The NP-hardness proof is done by a reduction from the NP-complete {\sc Exact Cover by 3-Sets} (XC3) problem; 
see Garey \& Johnson \cite{GaJo1979}.
An instance of XC3 consists of a ground set $X$ and a system $\cal T$ of 3-element subsets of $X$.
The problem is to decide whether there exists a partition of $X$ that only uses parts from $\cal T$.
We introduce for every element $x\in X$ three corresponding players $x$, $x'$ and $x''$ in a hedonic game.
The preferences of these players are as follows.
\begin{itemize}
\item The top choices of player $x$ are the triples $T\in\cal T$ with $x\in T$; he likes all of them equally much.
Strictly below these sets he (strictly) ranks the three coalitions $xx''\succ xx' \succ x$.
\item Player $x'$ only wants to be in three coalitions which he ranks $xx'\succ x'x'' \succ x'$.
\item Player $x''$ only wants to be in three coalitions which he ranks $x'x''\succ xx'' \succ x''$.
\end{itemize}
We claim that the constructed hedonic game has a core stable partition if and only if the XC3 instance 
has a feasible partition.
(Proof of if): Use the sets in the XC3 partition together with all sets $x'x''$.
The resulting partition is core stable, since every player $x\in X$ is in one of his most preferred coalitions.
(Proof of only if): If a partition for the hedonic game puts a player $x\in X$ into one of the three coalitions
$xx'',xx',x$, it cannot be core stable as the three players $x$, $x'$, $x''$ are essentially in the unstable
situation of Example~\ref{ex:3cycle}.
Hence every player $x$ must be placed into a group from $\cal T$, and a core stable partition induces a partition 
of $X$ with all parts from $\cal T$.
\xqed

\bigskip
Ballester \cite{Ballester2004} also extends his NP-completeness result to the case where all preference 
relations are strict; as a side result this yields NP-completeness of deciding the existence of a strictly
core stable partition under the individually rational encoding.
Next we turn to so-called additive hedonic games, which form a common generalization of many other hedonic 
games that will be discussed lateron in this survey.

\begin{Xdefinition}
\label{df:additive}
A hedonic game is \emph{additive}, if every player $i\in N$ has a real-valued preference function $v_i:N\to\rz$
so that $S\preceq_i T$ holds if and only if $\sum_{j\in S}v_i(j)\le \sum_{j\in T}v_i(j)$.
\end{Xdefinition}

In other words, in an additive hedonic game every player has a value for every other player, and the value of a 
coalition is simply the overall value of its members (from the view point of player $i$); hence every player can 
easily evaluate his profit from participating in a certain coalition.  
Additive preference structures allow a particularly natural and succinct representation, as they can be
specified by $n^2$ numbers for $n$ players.
Furthermore, additive preferences satisfy a number of desirable axiomatic properties; see Barber\`a, Bossert 
\& Pattanaik \cite{BaBoPa2004}.

\begin{Xexample}
\label{ex:additive}
Consider five players $a_0,a_1,a_2,a_3,a_4$ that are sitting (in this order) around a round table.
Every player $a_i$ assigns a value $v_i(a_{i+1})=1$ to the player to his right, a value $v_i(a_{i-1})=2$ to 
the player to his left, and a value $-4$ to the remaining two players (indices are taken modulo 5 throughout).
We claim that this additive hedonic game does not allow a core stable partition.

Indeed, any coalition of size three or more contains an unhappy player who rather would stay alone.
If a partition contains two single-player coalitions $\{a_i\}$ and $\{a_{i+1}\}$, then it would be blocked by
$\{a_i,a_{i+1}\}$.
In the only remaining case for a potentially core stable partition, there is one single-player coalition $\{a_i\}$ 
and two two-player coalitions $\{a_{i+1},a_{i+2}\}$ and $\{a_{i+3},a_{i+4}\}$; this partition is blocked by
$\{a_i,a_{i+1}\}$.
Hence there is no core stable partition (and there also is no strictly core stable partition).
\end{Xexample}

Sung \& Dimitrov \cite{SuDi2007} show that {\core} verification in additive hedonic games is strongly 
NP-complete; this also follows from Theorem~\ref{th:graph.e2}.
It took more time to fully understand the complexity of the {\core} existence problem for the additive case.
On the positive side, Observation~\ref{ob:basic.2} tells us that the problem is contained in $\Sigma_2^p$.
On the negative side, Sung \& Dimitrov \cite{SuDi2010} proved it to be NP-hard, and later Aziz, Brandt \& Seedig 
\cite{AzBrSe2011} extended the NP-hardness argument even to the symmetric case where $v_i(j)=v_j(i)$ holds 
for all players $i,j\in N$.
Finally Woeginger \cite{Woeginger2013} settled the problem by showing that is encapsulates the full difficulty 
of $\Sigma_2^p$.

\begin{Xtheorem}
\label{th:Woeginger}
(Woeginger \cite{Woeginger2013})~
In additive hedonic games, the {\core} existence problem is $\Sigma_2^p$-complete. 
\xqed
\end{Xtheorem}

Next, let us turn to \emph{strictly} core stable partitions in additive hedonic games.
The arguments of Sung \& Dimitrov \cite{SuDi2007} imply the NP-completeness of the verification question for this scenario.
Sung \& Dimitrov \cite{SuDi2010} prove NP-hardness of the existence question, but it seems very unlikely to me that
this problem could actually be contained in NP.
\begin{Xopen}
\label{open:strict}
{\opena}
Establish $\Sigma_2^p$-completeness of deciding whether a given additive hedonic game has
a \emph{strictly} core stable partition.
\end{Xopen}

\section{Preference structures from graphs}
\label{sec:graphs}
\nopagebreak
Dimitrov, Borm, Hendrickx \& Sung \cite{DiBoHeSu2006} study hedonic games where each player views every 
other player either as a friend or as an enemy: 
the players form the vertices of a directed graph $G=(N,A)$, an arc $(x,y)$ from player $x$ to player $y$ 
means that $x$ considers $y$ a friend, and the absence of such an arc means that $x$ considers $y$ an enemy.
We stress that here friendship is not a symmetric relation.
For a player $x\in N$, we denote by $F_x$ the set of his friends and by $E_x$ the set of his enemies.
Dimitrov \& al \cite{DiBoHeSu2006} introduce two concrete preference structures that we will dub
\emph{friend-oriented} and \emph{enemy-oriented}, respectively.

\begin{Xdefinition}
\label{df:graph}
Under friend-oriented preferences, player $x$ prefers coalition $T$ to coalition $S$ (that is, $S\preceq_x T$ 
with $x\in S$ and $x\in T$) 
\begin{itemize}
\item if $|S\cap F_x|<|T\cap F_x|$,~~ or
\item if $|S\cap F_x|=|T\cap F_x|$ and $|S\cap E_x|\ge|T\cap E_x|$.
\end{itemize}
Under enemy-oriented preferences, player $x$ prefers coalition $T$ to coalition $S$
\begin{itemize}
\item if $|S\cap E_x|>|T\cap E_x|$,~~ or
\item if $|S\cap E_x|=|T\cap E_x|$ and $|S\cap F_x|\le|T\cap F_x|$.
\end{itemize}
\end{Xdefinition}

Note that friend-oriented and enemy-oriented preferences both form special cases of additive preferences:
in the friend-oriented case, we set $v_x(y)=|N|$ if $x$ considers $y$ a friend and $v_x(y)=-1$ otherwise;
in the enemy-oriented  case, we set $v_x(y)=1$ if $x$ considers $y$ a friend and $v_x(y)=-|N|$ otherwise.
Although the definitions of these two preference structures are symmetric to each other, the two 
resulting classes of hedonic games do behave very differently from each other.
Let us start our discussion with the friend-oriented scenario.

\begin{Xtheorem}
\label{th:graph.f1}
(Dimitrov \& al \cite{DiBoHeSu2006})~
Under friend-oriented preferences, there always exists a core stable partition.
\end{Xtheorem}
\xproof
We use the strongly connected components $C_1,\ldots,C_k$ of the directed graph as partition.
Suppose for the sake of contradiction that there exists a blocking coalition $S$. 
Without much loss of generality we assume that $S$ intersects every component $C_i$ (as the
non-intersecting components can be safely ignored).
There exists a sink component $C_j$ without arcs to other components, and we distinguish two cases.
If $C_j\not\subseteq S$, then one of the vertices in $S\cap C_j$ has an arc into $C_j-S$; hence this 
vertex has fewer friends in $S$ than in $C_i$, and $S$ is not blocking.
If $C_j\subseteq S$, then every vertex in $C_j$ has the same number of friends in $C_j$ and in $S$,
but strictly more enemies in $S$; hence $S$ is not blocking.
\xqed

\bigskip
(A closer look at this proof shows that the strongly connected components actually form a \emph{strictly}
core stable partition.)
It is easy to see that in a core stable partition every coalition must be strongly connected, but there 
also exist examples with core stable partitions where every coalition is a proper subset of a strongly 
connected component:
assume that players $A_1$ and $A_2$ are mutual friends, that $B_1$ and $B_2$ are mutual friends, 
that $A_1$ is friendly towards $B_1$, and that $B_2$ is friendly towards $A_2$.
Then the partition $\{A_1,A_2\}$ and $\{B_1,B_2\}$ is core stable.

\begin{Xopen}
\label{open:graph.f}
{\opena}
Is there a polynomial time algorithm for the {\core} verification problem under friend-oriented preferences?
\end{Xopen}

We turn to the enemy-oriented scenario, and we start with a crucial observation:
if player $x$ considers player $y$ an enemy, then a core stable partition cannot place $x$ and $y$
together into the same coalition.
Such a coalition would violate individual rationality, as player $x$ would rather stay alone than
be together with $y$.
Consequently under the enemy-oriented scenario only mutual (symmetric) friendship relations matter,
and from now on we will assume that the underlying friendship graph $G$ actually is undirected.
Note furthermore that in a core stable partition every coalition induces a clique in $G$.

\begin{Xtheorem}
\label{th:graph.e1}
(Dimitrov \& al \cite{DiBoHeSu2006})~
Under enemy-oriented preferences, there always exists a core stable partition.
\end{Xtheorem}
\xproof
The game satisfies the common ranking property in Observation \ref{ob:basic.3}:
set $f(S)=|S|$ if $S$ induces a clique and $f(S)=0$ otherwise.
\xqed

\bigskip
Superficially, the results for friend-oriented preferences in Theorem~\ref{th:graph.f1} and for
enemy-oriented preferences in Theorem~\ref{th:graph.e1} have a very similar smell.
But the two problems differ a lot, if one actually wants to find such a core stable partition.
As the strongly connected components of a directed graphs can be determined in polynomial time (see for 
instance Cormen \& al \cite{CLRS}), in the friend-oriented scenario core stable partitions are easy to find.
On the other hand maximum cliques are NP-hard to find (Garey \& Johnson \cite{GaJo1979}), and every core stable 
partition in the enemy-oriented scenario must contain such a maximum clique; hence in the enemy-oriented scenario
core stable partitions are hard to find.

\begin{Xtheorem}
\label{th:graph.e2}
(Sung \& Dimitrov \cite{SuDi2007})~
Under enemy-oriented preferences, the {\core} verification problem is strongly NP-complete.
\end{Xtheorem}
\xproof
By Observation~\ref{ob:basic.1} the verification problem lies in NP.
NP-hardness is shown by a reduction from the NP-complete {\sc Maximum Clique} problem; see Garey \& 
Johnson \cite{GaJo1979}.
An instance of the clique problem consists of an undirected graph $G'=(V',E')$ and an integer bound $k$.
The problem is to decide whether the graph contains a clique on $k$ vertices.

We define a new graph $G=(V,E)$ by adding vertices and edges to $G'$: for every vertex $v\in V'$, we 
create $k-2$ new vertices that together with $v$ form a $(k-1)$-clique.
Finally we define the partition $\Pi$ of $V$ whose parts are exactly the vertex sets of these $(k-1)$-cliques.  
We claim that in the constructed hedonic game for $G$ there is a blocking set for partition $\Pi$
if and only if the graph $G'$ has a clique of size $k$.
(Proof of if): The clique of size $k$ forms a blocking set for $\Pi$.
(Proof of only if): If $G'$ has no clique of size $k$, the largest clique in graph $G$ has size $k-1$.
Hence $\Pi$ assigns every player to his most preferred coalition.
\xqed

\bigskip
Next let us discuss \emph{strictly} core stable partitions.
The proof of Theorem~\ref{th:graph.e2} also implies NP-completeness of the strict verification problem.
The strict existence problem seems to be fairly messy, and I would not be surprised if it turns out 
to be $\Sigma_2^p$-complete; note for instance that the path $P_n$ on $n\ge2$ vertices and the 
cycle $C_n$ on $n\ge4$ vertices allow a strictly core stable partition if and only if $n$ is even.

\begin{Xopen}
\label{open:graph.1}
{\opena}
Pinpoint the computational complexity of deciding whether a given hedonic game with enemy-oriented 
preferences has a \emph{strictly} core stable partition.
\end{Xopen}

Here is another variation.
For an undirected graph $G$ and a vertex $v\in G$, we let $\omega_G(v)$ denote the size of the largest
clique that contains $v$.
A partition of the vertices is called \emph{wonderfully stable}, if every vertex $v$ ends up in a
coalition of size $\omega_G(v)$.
In the enemy-oriented scenario, a wonderfully stable partition puts every player into his most preferred coalition.

\begin{Xopen}
\label{open:graph.2}
{\openb}
Pinpoint the computational complexity of deciding whether a given undirected graph has a wonderfully stable partition.
\end{Xopen}

The wonderfully stable partition problem is NP-hard, but perhaps unlikely to be contained in NP.
The problem is also unlikely to be $\Sigma_2^p$-complete, as it can be solved in polynomial time with
a logarithmic number of calls to an NP-oracle: 
the oracle algorithm first determines the value of $\sum_{v\in N}\omega_G(v)$ by a binary search;
every step in this binary search costs one call to the NP-oracle;
then the algorithm asks the NP-oracle whether there exists a partition into cliques that reaches this value.
This places the problem into the complexity class $\Theta_2^p$ which is believed to be a proper subset 
of $\Sigma_2^p$; see for instance Wagner \cite{Wagner1990} or Theorem {17.7} in \cite{PapaBook} for more 
information on this class.

\section{Anonymous preference structures}
\label{sec:anonym}
\nopagebreak
In a hedonic game with \emph{anonymous preferences}, every player is indifferent about coalitions of the same size.
Hence a player's preferences can be concisely specified by stating his ranking of coalition sizes.
A natural example for anonymous preferences is a chess club where all even group sizes should be fine, 
whereas odd groups sizes would prevent the players from splitting into chess-playing pairs.

In the verification problem, we search for a blocking coalition $S$ by checking the possible sizes $s=|S|$ one by one.
For a fixed size $s$, it is easy to decide whether there are $s$ players who would be happier in a coalition
of size $s$ than in their current coalitions.
Since the verification problem is polynomially solvable, Observation~\ref{ob:basic.2} yields that the
existence problem lies in NP.
Ballester \cite{Ballester2004} fully settles the complexity:

\begin{Xtheorem}
\label{th:anonymous}
(Ballester \cite{Ballester2004})~
Under anonymous preferences, the {\core} verification problem is polynomially solvable and
the {\core} existence problem is NP-complete.
\end{Xtheorem}

Darmann \& al \cite{Wine2012} consider a closely related scenario where every player has a primitive black-and-white 
view of the world: he (equally) likes some of the coalition sizes, and he (equally) hates the remaining ones.
The question is whether there exists a \emph{wonderfully stable} partition of the players, that is a partition so 
that each player likes the size of his coalition.

\begin{Xtheorem}
\label{th:Darmann}
(Darmann \& al \cite{Wine2012})~
Under anonymous black-and-white preferences, it is NP-complete to decide whether a hedonic game has a 
wonderfully stable partition.
\end{Xtheorem}
\xproof
We give a reduction from the {\sc Exact Cover by 3-Sets} (XC3) problem as defined in the proof of Theorem~\ref{th:Ballester}.
Let $T_1,\dots,T_m$ be an enumeration of the triples in $\cal T$, and define $\num(T_k)=4k$.
Here are our players:
\begin{itemize}
\item For every triple $T\in\cal T$, create $\num(T)-3$ players who like the two sizes $\num(T)-3$ and $\num(T)$
and hate all other sizes.
\item For every $x\in X$, create a single player $P(x)$ who only likes the sizes $\num(T)$ for the triples $T$
with $x\in T$ and hates all the other sizes.
\end{itemize}
It can be seen that the constructed hedonic game has a core stable partition if and only if the XC3 instance
has a feasible partition.
\xqed

\begin{Xopen}
\label{open:interval}
{\openb}
Consider the anonymous black-and-white hedonic game where every player $i$ likes the sizes $s$ in a certain 
interval $a_i\le s\le b_i$ and hates the remaining sizes.
Is there a polynomial time algorithm for finding a wonderfully stable partition?
\end{Xopen}

\section{Partition into pairs}
\label{sec:pair}
\nopagebreak
Throughout this section we only consider coalitions of size two.
Hence the preferences of a player can be specified by simply listing his ranking of the other players,
and the resulting hedonic games clearly are additive.
Since in this case the {\core} verification problem is straightforward (by searching through all pairs), 
we concentrate on the existence problem.

There are two basic variants that are known as the \emph{stable matching} (or \emph{stable marriage}) 
problem and as the \emph{stable roommate} problem.
The stable matching problem has a bipartite structure: there are $n$ male and $n$ female players, and 
the only feasible pairs are man-woman couples.
The stable roommate problem has a non-bipartite structure: there are $2n$ unisex players, and every 
possible pair is feasible.

\subsection{Stable matchings}
\label{ssec:pair.1}
\nopagebreak
The stable matching problem was introduced in the seminal paper by Gale \& Shapley \cite{GaSh1962},
one of the most cited papers in computational social choice.
A matching $\mu$ pairs the men with the women; the partner of man $M$ in the matching is denoted 
$\mu(M)$ and the partner of woman $W$ is denoted $\mu(W)$.
A man $M$ and a woman $W$ form a \emph{blocking pair} $(M,W)$ for matching $\mu$, if $M$ prefers $W$ to 
his woman $\mu(M)$ and if simultaneously $W$ prefers $M$ to her man $\mu(W)$.

Perhaps the most natural approach to stable matching would be the following iterative improvement procedure:
Start with an arbitrary matching, and then iteratively find a blocking pair $(M,W)$ and improve the situation
by replacing the two pairs $(M,\mu(M))$ and $(\mu(W),W)$ by the new pairs $(M,W)$ and $(\mu(W),\mu(M))$.
The following example demonstrates that this idea may fail horribly.

\begin{Xexample}
\label{ex:Tamura}
(Tamura \cite{Tamura1993})~
There are four men $A,B,C,D$ and four women $W,X,Y,Z$ with the following preference lists:
\\[1.0ex]\indent
\begin{tabular}{ll@{\qquad\qquad}ll}
$A$: &$X>Z>W>Y$   &$W$: &$A>C>B>D$ \\
$B$: &$Y>W>X>Z$   &$X$: &$B>D>C>A$ \\
$C$: &$Z>X>Y>W$   &$Y$: &$C>A>D>B$ \\
$D$: &$W>Y>Z>X$   &$Z$: &$D>B>A>C$
\end{tabular}
\\[1.0ex]
Assume that the iterative improvement procedure picks $AW, BX, CZ, DY$ as its starting point.
The following lines show the dynamics of the resulting process. 
\begin{enumerate}
\item~ $\A W, BX, C\Z, DY$ \qquad $A$ and $Z$ are blocking
\item~ $AZ, BX, \C W, D\Y$ \qquad $C$ and $Y$ are blocking 
\item~ $AZ, \B X, CY, D\W$ \qquad $B$ and $W$ are blocking 
\item~ $A\Z, BW, CY, \D X$ \qquad $D$ and $Z$ are blocking 
\item~ $A\X, BW, \C Y, DZ$ \qquad $C$ and $X$ are blocking 
\item~ $\A Y, B\W, CX, DZ$ \qquad $A$ and $W$ are blocking
\item~ $AW, B\Y, CX, \D Z$ \qquad $D$ and $Y$ are blocking 
\item~ $AW, \B Z, C\X, DY$ \qquad $B$ and $X$ are blocking 
\end{enumerate}
The last improvement yields the matching $AW, BX, CZ, DY$, so that we are back at our starting point.
The process is cycling and will never terminate!
\end{Xexample}

Let us take a closer look at the instance in Example~\ref{ex:Tamura}. 
There are 24 possible matchings, five of which are actually stable.
If we start the iterative improvement procedure from one of the three matchings $AY, BZ, CW, DX$ or
$AW, BZ, CY, DX$ or $AY, BX, CW, DZ$, then the process will eventually reach the stable matching 
$AW, BX, CY, DZ$ and terminate.
But if we start the iterative improvement procedure from any of the remaining 16 matchings, then the procedure 
will cycle and does not terminate.

Fortunately, there exist better approaches for the stable matching problem: it can be shown that there 
\emph{always} exists a stable solution, which furthermore can be computed in polynomial time by the 
celebrated Gale-Shapley algorithm \cite{GaSh1962}.
As the books by Knuth \cite{KnuthBook}, Gusfield \& Irving \cite{IrvingBook}, and Roth \& Sotomayor 
\cite{RothBook} comprehensively analyze this algorithm and extensively cover the combinatorial facets of
the problem, we only formulate a summarizing theorem.

\begin{Xtheorem}
\label{th:GS}
(Gale \& Shapley \cite{GaSh1962})~
If all preferences are strict, a stable matching always exists and can be found in polynomial time.
\xqed
\end{Xtheorem}

We now would like to spend some lines on several closely related variants.
Theorem~\ref{th:GS} assumes that every player has a strict preference ranking of the other players.
Allowing \emph{ties} in the preference relations does not change much: a stable matching always exists 
and can be found in polynomial time (by breaking ties arbitrarily and then applying Gale-Shapley); 
see for instance Irving \cite{Irving1994}.  

Allowing \emph{incomplete preference lists} (but still forbidding ties) changes the situation a little bit.
Now every player can exclude some other players with whom he does not want to be matched (formally this
can be done by ranking the unwanted coalitions below the coalition where he stays alone).
It turns out that also for this case a stable matching always exists and can be found in polynomial time 
by a slight modification of the Gale-Shapley algorithm.
However a stable matching is not necessarily perfect: it will consist of some pairs and of some isolated singletons.
Interestingly \emph{every} stable matching has the same set of men and women paired up and the same set of 
men and women as singletons; see Gale \& Sotomayor \cite{GaSo1985}.

Simultaneously allowing both \emph{incomplete preference lists} and \emph{ties} messes things up a lot.
A stable matching always exists and can be found in polynomial time, but the same instance can have very
different stable matchings with varying numbers of pairs.
Deciding whether there is a perfect stable matching (which pairs up all the players) is NP-complete; see
Manlove, Irving, Iwama, Miyazaki \& Morita \cite{MaIrIwMiMo2002}.
In fact this perfect stable matching variant is NP-complete even if the preference list of every player
lists only three acceptable partners; see Irving, Manlove \& O'Malley \cite{IrMaMa2009}.

Finally Irving \& Leather \cite{IrLe1986} have shown that counting the number of stable matchings
(in the classical version without ties and without incomplete preference lists) is \#P-complete
Chebolu, Goldberg \& Martin \cite{ChGoMa2010} indicate that even approximate counting should be difficult.

\subsection{Stable roommates}
\label{ssec:pair.2}
\nopagebreak
The stable roommate problem is the non-bipartite unisex version of the stable matching problem.
The following example demonstrates that there are roommate instances without stable matching
(note the structural similarity between Example~\ref{ex:3cycle} and Example~\ref{ex:roommate}).

\begin{Xexample}
\label{ex:roommate}
Consider a situation with four players $A,B,C,D$ that have the following preferences:
player $A$ prefers $B>C>D$; 
player $B$ prefers $C>A>D$;
player $C$ prefers $A>B>D$;
and player $D$ prefers $A>B>C$.
Note that none of $A,B,C$ wants to play with the unpopular dummy player $D$.

The 
matching $\{AB,CD\}$ is blocked by $BC$, and 
matching $\{AC,BD\}$ is blocked by $AB$, and
matching $\{BC,AD\}$ is blocked by $AC$.
Hence there is no core stable partition. 
\end{Xexample}

A milestone paper by Irving \cite{Irving1985} characterizes the roommate instances with core stable matchings.
\begin{Xtheorem}
\label{th:Irving}
(Irving \cite{Irving1985})~
For the stable roommate problem with strict preferences, the existence of a stable matching 
can be decided in polynomial time.
\xqed
\end{Xtheorem}

If we allow \emph{incomplete preference lists} (but still forbid ties), a minor modification of Irving's
algorithm \cite{Irving1985} solves the stable roommate problem in polynomial time.
If we allow \emph{ties} in the preference relations, the stable roommate problem becomes NP-complete;
see Ronn \cite{Ronn1990} and Irving \& Manlove \cite{IrMa2002}.

Arkin, Bae, Efrat, Okamoto, Mitchell \& Polishchuk \cite{Arkin2009} discuss a metric variant of the
stable roommate problem where every player is a point in a metric space with distance function $|\cdot|$.
Player $P$ prefers being with player $X$ to being with player $Y$ if and only if $|PX|\le|PY|$.
This special case always has a stable matching, as it satisfies the common ranking property of 
Observation \ref{ob:basic.3}: just set $f(XY)=-|XY|$ for coalitions $XY$ of size two \cite{Arkin2009}.

\section{Partition into triples}
\label{sec:triple}
\nopagebreak
Generalizations of the classical Gale-Shapley stable matching problem (with men and women as the two genders) 
to three genders (men, women, dogs) usually are very messy.
Alkan \cite{Alkan1988} seems to have been the first to publish a result on this 3-gender variant,
by constructing a concrete example that does not allow a core stable matching. 
The preferences in Alkan's example are additively separable, and there are $n=3$ men, women and dogs.
Ng \& Hirschberg \cite{NgHi1991} exhibit an even smaller bad instance with $n=2$:

\begin{Xexample}
\label{ex:NH}
(Ng \& Hirschberg \cite{NgHi1991})~
Consider two men $M_1,M_2$, two women $W_1,W_2$ and two dogs $D_1,D_2$ that have the following 
preferences over the triples:
\begin{itemize}
\item[~]$M_1$:~~ $M_1W_1D_2 ~>~ M_1W_1D_1 ~>~ M_1W_2D_2 ~>~ M_1W_2D_1$
\item[~]$M_2$:~~ $M_2W_2D_2 ~>~ M_2W_1D_1 ~>~ M_2W_2D_1 ~>~ M_2W_1D_2$
\item[~]$W_1$:~~ $M_2W_1D_1 ~>~ M_1W_1D_2 ~>~ M_1W_1D_1 ~>~ M_2W_1D_2$
\item[~]$W_2$:~~ $M_2W_2D_1 ~>~ M_1W_2D_1 ~>~ M_2W_2D_2 ~>~ M_1W_2D_2$
\item[~]$D_1$:~~ $M_1W_2D_1 ~>~ M_1W_1D_1 ~>~ M_2W_1D_1 ~>~ M_2W_2D_1$
\item[~]$D_2$:~~ $M_1W_1D_2 ~>~ M_2W_2D_2 ~>~ M_1W_2D_2 ~>~ M_2W_1D_2$
\end{itemize}
There are only four possible partitions into two disjoint triples:
\begin{itemize}
\item[~]The partition $\{M_1W_1D_1,~M_2W_2D_2\}$ is blocked by $M_1W_1D_2$.
\item[~]The partition $\{M_1W_1D_2,~M_2W_2D_1\}$ is blocked by $M_2W_1D_1$.
\item[~]The partition $\{M_1W_2D_1,~M_2W_1D_2\}$ is blocked by $M_1W_1D_2$.
\item[~]The partition $\{M_1W_2D_2,~M_2W_1D_1\}$ is blocked by $M_2W_2D_2$.
\end{itemize}
Hence there exists no core stable matching.
\end{Xexample}

Ng \& Hirschberg \cite{NgHi1991} also establish the NP-completeness of deciding the existence of a core 
stable matching; this result has also been derived by Subramanian \cite{Subramanian1994} 
(independently and by a very different approach). 

Donald Knuth \cite{KnuthBook} proposes the 3-gender stable matching variant with so-called
\emph{cyclic preferences}: every man $M$ has a strict ordering of the women in the instance, every 
woman $W$ has a strict ordering of the dogs, and every dog $D$ has a strict ordering of the men.
A triple $MWD$ is blocking for a given current partition into triples, if man $M$ prefers $W$ to his 
currently assigned woman, if woman $W$ prefers $D$ to her currently assigned dog, and if dog $D$ prefers 
$M$ to its currently assigned man.
Boros, Gurvich, Jaslar \& Krasner \cite{BoGuJaKr2004} prove by case distinctions that every cyclic instance
with $n=3$ has a core stable matching, and Eriksson, Sj{\"o}strand \& Strimling \cite{ErSjSt2006} extend this
positive result to $n=4$.
The approaches in \cite{BoGuJaKr2004,ErSjSt2006} are quite technical and involve much case analysis, and
they do not seem to generalize to larger values of $n$.

\begin{Xopen}
\label{open:cyclic}
{\openc}
Prove that every instance of the 3-gender stable matching problem with cyclic preferences has a stable solution.
\end{Xopen}

Bir\'o \& McDermid \cite{BiDe2010} consider the case of cyclic preferences with unacceptable partners:
every man finds certain women unacceptable, every woman hates certain dogs, and every dog dislikes certain men.
Under this scenario there exist instances without stable solution, and deciding the existence of a stable 
solution is NP-complete.

Danilov \cite{Danilov2003} discusses a related (but much easier) special case where every man primarily cares 
about women and where every woman primarily cares about men (and where the preferences of the dogs are arbitrary).
This special case always has a stable matching.
In a first step, we find a stable matching for the underlying 2-gender instance that consists of men and women.
The preferences of men and women in this 2-gender instance are their primary rankings in the 3-gender instance.  
In the second step, we find a stable matching for the 2-gender instance with dogs on the one side and on
the other side the man-woman pairs from the first step.
The preferences of the dogs on man-woman pairs are copied from the 3-gender instance.
The preferences of the man-woman pairs on the dogs are always fixed according to the man in the pair:
the pair $MW$ prefers dog $D$ to dog $D'$, if and only if in the 3-gender instance man $M$ prefers triple
$MWD$ to triple $MWD'$.
Everything else follows from the Gale-Shapley Theorem~\ref{th:GS}.

In the 3-dimensional roommate problem all players have the same gender and every triple is a potential coalition.
Ng \& Hirschberg \cite{NgHi1991} establish NP-completeness of deciding the existence of a core stable matching
in the 3-dimensional roommate problem.
Arkin, Bae, Efrat, Okamoto, Mitchell \& Polishchuk \cite{Arkin2009} discuss the following geometric variant 
of the 3-dimensional roommate problem in the Euclidean plane with distance function $|\cdot|$:
every player is a point in the Euclidean plane, and player $P$ prefers triple $PX_1X_2$ to triple $PY_1Y_2$ 
if and only if $|PX_1|+|PX_2|<|PY_1|+|PY_2|$.
(Note that here the matching problem is 3-dimensional, whereas the underlying geometric space is 2-dimensional.
Note furthermore that the preference structure is additive.)
Arkin \& al \cite{Arkin2009} exhibit a highly structured instance that does not possess a core stable matching; 
the computational complexity of this special case however remains open.

\begin{Xopen}
\label{open:geometric}
{\openb}
Settle the complexity of the Euclidean 3-dimensional roommate problem as described in the preceding paragraph.
\end{Xopen}

\section{Preference structures from maxima and minima}
\label{sec:minmax}
\nopagebreak
Cechl\'arov\'a \& Romero-Medina \cite{CeRo2001} investigate hedonic games where every player ranks
his coalitions according to the most or least attractive member of the coalition.
Similarly as in the additive games in Definition~\ref{df:additive}, every player $i\in N$ has a 
real-valued function $v_i:N\to\rz$ that measures his addiction to each of the other players.
For a coalition $S$, we define $\vmax_i(S)=\max_{j\in S}v_i(j)$ as player $i$'s addiction to the 
best member of $S$.

\begin{Xdefinition}
\label{df:minmax}
Under max-preferences, player $i$ prefers coalition $T$ to coalition $S$ (that is, $S\preceq_i T$
with $i\in S$ and $x\in T$)
\begin{itemize}
\item if $\vmax_i(S)<\vmax_i(T)$,~~ or
\item if $\vmax_i(S)=\vmax_i(T)$ and $|S|\ge|T|$.
\end{itemize}
\end{Xdefinition}

An important special case of this scenario are max-preferences \emph{without ties}; this means that for 
distinct players $a$ and $b$ the values $v_i(a)$ and $v_i(b)$ assigned to them by player~$i$ are always distinct.

\begin{Xtheorem}
\label{th:max.v}
(Cechl\'arov\'a \& Hajdukov\'a \cite{CeHa2002})~
Under max-preferences, the {\core} verification problem is polynomially solvable.
\end{Xtheorem}
\xproof
How would we verify the core stablitiy of a given partition $\Pi$?
The main idea is to check step by step for $k=1,2,\ldots,|N|$ whether there exists a blocking 
coalition $S$ \emph{of size at most $k$}.
For checking a concrete value $k$, we construct an auxiliary directed graph $G_k$ on the vertex set $N$; 
an arc $i\to j$ means that player $i$ strictly prefers every coalition $S$ with $j\in S$ and $|S|\le k$ 
to his current coalition $\Pi(i)$.
Formally the graph $G_k$ contains the arc $i\to j$ if $\vmax_i(\Pi(i))<v_i(j)$ holds, 
or if $\vmax_i(\Pi(i))=v_i(j)$ and $|S|>k$.

If graph $G_k$ contains a directed cycle of length at most $k$, the corresponding vertices form a blocking
coalition of size at most $k$.
Vice versa, a blocking coalition of size at most $k$ induces a subgraph in $G_k$ with a cycle of length 
at most $k$.
The shortest cycle in a directed graph can be found in polynomial time; see for instance Cormen \& al \cite{CLRS}.
\xqed

\begin{Xtheorem}
\label{th:max.e}
(Cechl\'arov\'a \& Hajdukov\'a \cite{CeHa2002})~
Under max-preferences without ties, there always exists a core stable partition.
\end{Xtheorem}
\xproof
Make every player $i$ point at the player whom he likes most.
Then the underlying directed graph contains a cycle.
Pick the players along such a cycle as coalition $S$, and repeat this procedure for the remaining graph.
\xqed

\bigskip
The algorithm in the proof of Theorem~\ref{th:max.e} is essentially the famous top-trading-cycle 
algorithm of David Gale for the house swapping game (see for instance \cite{ShSc1974}).
On the negative side Cechl\'arov\'a \& Hajdukov\'a \cite{CeHa2002} prove that under max-preferences 
\emph{with ties} the {\core} existence problem is NP-complete.

In a closely related line of research Cechl\'arov\'a \& Hajdukov\'a \cite{CeHa2004} investigate
\emph{min-preferences} where every player ranks his coalitions according to the \emph{least} attractive 
member in the coalition.
Under this scenario unstable partitions always have small blocking sets of size at most~2, so that the
{\core} verification problem is straightforward to solve.
Furthermore stable partitions always consist of small coalitions of size at most~3. 
For min-preferences \emph{without ties}, Cechl\'arov\'a \& Hajdukov\'a \cite{CeHa2004} design a 
modification of Irving's roommate algorithm (Theorem~\ref{th:Irving}) that solves the {\core} existence 
problem in polynomial time.
They also show that for min-preferences \emph{with ties} the existence problem is NP-complete.


\end{document}